\documentclass[preprint,amssymb,amsmath]{revtex4-1}

\usepackage{graphicx}%

\usepackage{mathtools}   
\usepackage{cases}

\begin{document}

\title{Analytic Structure of the S-Matrix 
\\
for 
Singular Quantum Mechanics}

\author{
 Horacio E. Camblong,$^{1}$ 
Luis N. Epele,$^2$ Huner Fanchiotti,$^2$
 and Carlos A. Garc\'{\i}a Canal$^2$}
 
\affiliation{
$^{1}$ Department of Physics, 
University of San Francisco, San Francisco, California 94117-1080, USA \\
$^{2}$  Laboratorio de F\'{\i}sica Te\'{o}rica,
 Departamento de F\'{\i}sica, IFLP, CONICET,
Facultad de Ciencias Exactas,
Universidad Nacional de La Plata,
 C.C. 67--1900 La Plata, Argentina
}

\begin{abstract}
The analytic structure 
of the S-matrix of singular quantum mechanics 
is examined within a multichannel framework,
with primary focus on its dependence 
with respect to a parameter ($\Omega$) that determines the boundary conditions.
Specifically, a characterization is given in terms of salient mathematical and
physical properties governing its behavior.
These properties involve
unitarity and associated current-conserving Wronskian relations, time-reversal invariance, and
Blaschke factorization.
The approach leads to an interpretation of effective nonunitary solutions in singular quantum mechanics 
and their determination from
the unitary family.

\end{abstract}


\maketitle

\section{Introduction}
\label{sec:introduction}

Singular potentials are known to pose notoriously subtle difficulties that call for an extension
of the rules of ordinary quantum mechanics.
Some of the outstanding problematic features
of singular quantum mechanics (SQM) were addressed
 in the pioneering work of Ref.~\cite{cas:50} and
in the early literature reviewed in Refs.~\cite{spector_RMP} and \cite{Newton}.
These features can be ultimately traced to an indeterminacy in the boundary conditions~\cite{landau:77}.
More recent general frameworks were proposed in Ref.~\cite{Esposito} using 
polydromy properties~\cite{Stroffolini}, 
and in  Ref.~\cite{Smatrix-singularQM} within a general multichannel framework.
In addition, significant progress has been made for specific problems related to the renormalization 
of the inverse square potential (ISP)~\cite{Gupta-1993,HEC-LP1,HEC-LP2,HEC:CQM-renormalization}
and contact interactions~\cite{Jackiw-1991-Beg}, 
their associated conformal symmetry~\cite{Alfaro-Fubini-Furlan-1976,Jackiw_SO(21)_1,Jackiw_SO(21)_2}
and anomaly~\cite{HEC-CQManomaly1,HEC-CQManomaly2},
and other cases of SQM renormalization~\cite{Beane:2000wh}. 

In this paper, we reexamine the generic form of the S-matrix for SQM and
shed light on the existence of an intriguing connection between the unitary and nonunitary solutions.
A particular form of this connection can be found in the pioneering work of
 Ref.~\cite{radin:75}, where it was pointed out that,
 while the basic results of Ref.~\cite{cas:50} focus on the unitary but nonunique solutions of SQM,
 the path-integral treatment of Ref.~\cite{nel:64} for the inverse square potential (ISP)
 generates a distinct and unique nonunitary solution with perfect absorption. 
Thus, in Ref.~\cite{radin:75}, an average expression 
at the level of Green's functions 
is used to relate particular cases of unitary and nonunitary solutions.
Subsequently, it was pointed out~\cite{bawin_coon_wire} 
that a similar result can be obtained
for the S-matrix of the two-dimensional ISP:  an 
average of unitary solutions coincides with the solution for perfect absorption. 

In essence, we display the existence of a network of relations supported on
current-conservation equations or concomitant unitarity of an appropriately defined S-matrix.
We implement our proposal within our recently developed general multichannel 
framework~\cite{Smatrix-singularQM}
that not only includes the ISP in any number of dimensions but also applies to more  singular potentials. 
Thus, we address broad analytic properties of the S-matrix that can be derived from general principles---determining, 
among other things, the functional Blaschke-type form of the 
S-matrix~\cite{Blaschke1,Blaschke2, Garnett:1981}, 
in addition to other related mathematical properties.

The remainder of this paper is organized as follows.
In Sec~\ref{sec:SQM-multichannel-setup} 
we outline the basic definitions and setup of the multichannel SQM framework.
Section~\ref{sec:S-matrix-structure}
is the core of the paper, with an examination of the analytic properties of the S-matrix
vis-\`{a}-vis its boundary value indeterminacy.
Section~\ref{sec:unitary-nonunitary_connection}
explores the unitary-nonunitary connection arising from the S-matrix structure.
The conclusions are summarized in 
Sec~\ref{sec:conclusions}.

\vspace*{-0.075in}

\section{Singular Quantum Mechanics (SQM): Multichannel  Framework and Setup of the Problem}
\label{sec:SQM-multichannel-setup}

Following Ref.~\cite{Smatrix-singularQM}, 
 we consider a quantum-mechanical problem or equivalent system described by a 
Schr\"{o}dinger-type equation, i.e.,
 the normal invariant form of the generic linear second-order differential equation that can be obtained
 via a Liouville transformation~\cite{Forsyth}:
\begin{equation}
\left[
\frac{d^{2}}{dr^{2}} 
\,
+ J(r)
\right]  
u (r) = 0
\;  .
\label{eq:singular-QM_radial}
\end{equation}
In Eq.~(\ref{eq:singular-QM_radial}), for a broad range of physical 
applications, one can rewrite the normal invariant~\cite{Forsyth} as 
$J(r)
=
 k^{2} 
-
V( r )
-
 \left[\left( l + \nu \right)^{2} - 1/4 \right]/r^{2}
$, 
with 
parameters $l$ and $\nu$  (usually associated with angular momentum and spatial dimensionality)
 to be adjusted for each physical or mathematical application (details can be found in 
 Refs.~\cite{Smatrix-singularQM} and
 \cite{HEC:CQM-renormalization}, where a class of
 anisotropic potentials are also subsumed in this formulation).

 In addition to an irregular singularity at infinity, 
 we assume the existence of a singular point at a finite location that we take to be $r=0$, 
 with behavior 
 \begin{equation}
 J( r ) 
 \sim 
 -
 V( r ) 
 \sim 
   \lambda
\,
 r^{-p}
 \label{eq:potential-near-singularity}
 \end{equation}
for $r \sim 0$, where
\begin{itemize}
\item
$p=2$, with a regular singularity.
This marginally singular case is known as the ISP,
defining long-range conformal quantum mechanics (CQM).

\item $p>2$, with an irregular singularity. This properly singular case 
yields the family of strongly singular power-law potentials.

\end{itemize}
Thus, dominant contribution~(\ref{eq:potential-near-singularity})
 in the neighborhood of the origin 
defines
the all-important  power-law class of potentials in SQM.
Moreover, with minor adjustments,
our problem could be
 generalized to include additional singular points, as well as logarithmic behavior near the 
singularities.

Let us consider the  set of solutions
${\mathcal B}_{\rm sing}
=
\biggl\{
u_{+}(r) , u_{-}(r)
\biggr\}
\;  
\label{eq:singularity-basis}
$
with {\em outgoing/ingoing wave-like behavior
near  
the singularity at $r=0$\/}.
This solution set serves as a fundamental basis for the two-dimensional solution space.
Thus, it conveys information regarding the singular point
as a generalized form of the solutions proposed in Ref.~\cite{Vogt-Wannier} 
for the inverse quartic potential.
Some explicit expressions are shown below.
 
In addition, the singular point at infinity 
involves another set of solutions
with
{\em outgoing/ingoing wave-like behavior
 as $r \sim \infty$\/}:
$
{\mathcal B}_{\rm asymp}
=
\biggl\{
u_{1}(r) , u_{2}(r)
\biggr\}
$, 
and which are 
 motivated by the
determination of physical observables via 
the {\em asymptotic S-matrix\/} $ S_{\rm asymp} $
(see below).
Explicitly,
\begin{equation}
u_{1,2} (r)
\stackrel{( r \rightarrow \infty )}{\sim}
\frac{1}{ \sqrt{k} }
\,
e^{\mp i \pi/4}
\,
e^{\pm i kr} 
\; ,
\label{eq:asymptotic-wave-normalization}
\end{equation}
where: (i)  the chosen phases are adopted for comparison with 
asymptotic expansions of 
Hankel functions; (ii) the normalization is enforced with the WKB amplitude factor $k^{-1/2}$,
which is re-evaluated in Sec.~\ref{sec:S-matrix-structure}
 from Wronskian properties.
Correspondingly, the linear relation between
the two bases, with the associated resolutions of $u(r)$,
\begin{equation}
u  (r) 
=
C^{ \mbox{\tiny $ (+)$} } 
u_{+}(r) 
+  
C^{ \mbox{\tiny $ (-)$} } 
u_{-} (r)
=
C^{ \mbox{\tiny $ (1)$} } 
u_{1}(r) 
+ 
C^{ \mbox{\tiny $ (2)$} } 
  u_{2} (r)
\; ,
\label{eq:wave-function_bases}
\end{equation}
provide a formal solution to the most general problem of SQM with one {\em finite singular point\/}, as sketched in
Fig.~1.
In summary, the connection between the bases
${\mathcal B}_{\rm sing}$
and
${\mathcal B}_{\rm asymp}$
constitutes a two-channel framework,
where each ``channel'' is associated with a singular point  (with one finite singular point and
the second singular point located at infinity)---generalizations to multiple singularities are
possible as a multichannel setup~\cite{Smatrix-singularQM},
as will be further discussed elsewhere.

In particular, 
it is convenient to rewrite the distinct  sides 
of Eq.~(\ref{eq:wave-function_bases})
as
\begin{equation}
u  (r) \propto 
\Omega 
\, u_{+}(r) +  u_{-} (r)
\; 
\label{eq:wave-function_near-origin-basis}
\end{equation}
and
\begin{equation}
u  (r) \propto 
\hat{S}_{\rm asymp} 
\, 
u_{1} (r) +  u_{2} (r) 
\;  .
\label{eq:wave-function_asymptotic-basis}
\end{equation}
Here, with the use of the proportionality symbol,
the ratio
\begin{equation}
\Omega = 
\frac{
C^{ \mbox{\tiny $ (+)$} } 
}{
C^{ \mbox{\tiny $ (-)$} } 
}
\; 
\label{eq:singularity-parameter}
\end{equation}
provides 
 a ``singularity parameter'' 
 related to the indeterminacy of the boundary conditions at the finite singular point. 
 Thus, $\Omega$
 specifies an auxiliary ``boundary condition,''
i.e., it gauges the relative probability amplitudes of 
outgoing (emission) to ingoing (absorption) waves
in the  neighborhood of $r=0$.
Similarly,
we fix the normalization for the point at infinity with
\begin{equation}
S_{\rm asymp} 
= 
e^{i \pi \left( l+\nu \right)} 
\, 
\hat{S}_{\rm asymp} 
\; , 
\label{eq:S-matrix_and_reduced-S-matrix}
\end{equation}
in terms of 
the {\em reduced matrix elements\/}
$\hat{S}_{\rm asymp}$ 
and an $l$- and $d$-dependent phase factor.
 Equations~(\ref{eq:wave-function_asymptotic-basis}) and (\ref{eq:S-matrix_and_reduced-S-matrix})
yield the S-matrix from which the physical observables are extracted.
In addition, by the way they are constructed as outgoing/ingoing waves,
the basis functions satisfy
\begin{align}
 [u_{1} (r)]^{*} & =  u_{2} (r) 
 \label{eq:conjugate-basis-12-relation}
\; ,
\\
 [u_{+} (r)]^{*} & =  u_{-} (r) 
 \label{eq:conjugate-basis-pm-relation}
 \; ,
\end{align}
for $r \in \mathbb{R}$.

For the sake of completeness, we provide explicit expressions for the dominant behavior of the ``singularity
basis''~\cite{Smatrix-singularQM},
\begin{numcases}
  {
  u_{\pm} (r)   
   \stackrel{( r \rightarrow 0 )}{\sim} 
    }
\frac{ r^{p/4} }{{\lambda}^{1/4}}
\,
\exp \!
\left[
\mp 2 \, i \, {\lambda}^{1/2}
\,
 \frac{ 
 r^{ - \left(  p /2 - 1 \right) } 
}{
\left( p - 2 \right)}
\right]
&
\text{  for  $p>2$}
\; ,
\label{eq:Wannier_waves_p>2} 
\\
 {\displaystyle
\sqrt{ \frac{ r }{\Theta } }
\,
\exp \!
\left[ 
\pm i \, \Theta
\,
\ln \left( \mu r \right)
\right]
}
=
\sqrt{ \frac{ r }{\Theta } }
\;
\left( \, \mu \, r \, \right)^{ \pm  i \, \Theta}
&
\text{  for  $p=2$} 
\; .
  \label{eq:Wannier_waves_p=2}
 \end{numcases}
Here, the normalization is also enforced with the WKB amplitude factor $(k_{\rm WKB})^{-1/2}$,
see again further details in Sec.~\ref{sec:S-matrix-structure}.
Moreover, the arbitrary floating inverse length $\mu$ for the regular singular case $p=2$, which
is mandatory by its asymptotic conformal invariance, arises from the integration constant 
in the WKB solution or via dimensional homogeneity in the associated 
Cauchy-Euler differential equation.
However, for the irregular singular case $p>2$,
the integration constant in the exponent only appears at a higher order in the asymptotic expansion 
with respect to $1/r$; specifically,
the Bessel-function solutions of Eq.~(\ref{eq:singular-QM_radial})
as  $r \rightarrow 0$, 
$u / \sqrt{r} \propto H^{(1,2)}_{-1/n}  \left( - 2 \sqrt{\lambda} \, r^{-n/2}/n \right)$, where $n= p-2>0$, combined with the 
asymptotics of Hankel functions, provide the correct 
outgoing/incoming behavior
of Eq.~(\ref{eq:Wannier_waves_p>2}).
In addition, the conformal case
 $p=2$ includes the Langer correction~\cite{Langer}
 corresponding to the critical coupling $\lambda =1/4$, with
shifted square-root coupling constant 
$
 \Theta^{2}
 \equiv
\lambda - 1/4 
$; in this case, for
 particular instances of nonrelativistic quantum mechanics,
  the angular momentum
  is merged with the marginally singular $p=2$ term at the same order, leading to 
  an effective  interaction coupling---but this does not occur 
 for quantum fields in black hole backgrounds and other relativistic 
 applications~\cite{Smatrix-singularQM, HEC-CQManomaly1,HEC-CQManomaly2,HEC:CQM-renormalization}.
 Parenthetically,
when  the potential has a long-range tail for $r \sim \infty$ given by
$V (  r )  \sim - \lambda r^{-\delta}$,
the asymptotic behavior involves an
 extra phase in the form
  $\sqrt{k} \, u_{1,2} (r)
 \stackrel{( r \rightarrow \infty)}{\sim} 
\,
e^{\mp i \pi/4}
\,
e^{\pm i kr} 
\,
e^{\pm i \lambda r^{1-\delta}/2k (1 -\delta)}
$
which can be similarly derived by WKB integration 
for the irregular singular point at infinity~\cite{Smatrix-singularQM}.

\section{Analytic Structure of the S-matrix}
\label{sec:S-matrix-structure}

For the remainder of this paper,
we will write the reduced asymptotic S-matrix,
defined by Eq.~(\ref{eq:wave-function_asymptotic-basis}),
as
$\hat{S}   \equiv \hat{S}_{\rm asymp}  $.
The S-matrix of physical interest follows from Eq.~(\ref{eq:S-matrix_and_reduced-S-matrix}).
From Sec.~\ref{sec:SQM-multichannel-setup},
the existence of a complex function 
\begin{displaymath}
\hat{S}
=
\hat{S} (\Omega)
\end{displaymath}
of the variable $\Omega$
relies on the singular nature of the potential
via Eqs.~(\ref{eq:wave-function_near-origin-basis}) and (\ref{eq:wave-function_asymptotic-basis}).
 Specifically, the boundary condition indeterminacy at the finite singular point (e.g., the origin)
 is described by the arbitrary parameter $\Omega$  that represents 
 the ratio of the amplitude coefficients for the required outgoing and ingoing waves
 at the singularity.
 By the nature of the solutions, $\Omega$ is generically a complex number.
 
In addition, 
$ \hat{S} (\Omega)$ 
is a
{\em meromorphic function\/},
 as follows  constructively
 from the definition of the ``parameters'' 
  $\hat{S}$ 
  and $\Omega$
in the scattering process.
In effect,
the linear relationship between the coefficients 
$(C^{ \mbox{\tiny $ (+)$} } , C^{ \mbox{\tiny $ (-)$}}) $
and
$(C^{ \mbox{\tiny $ (1)$} } , C^{ \mbox{\tiny $ (2)$}}) $, or between the corresponding bases,
as displayed by Eq.~(\ref{eq:wave-function_bases}),
implies a fractional linear transformation for the relationship between 
$\hat{S}$ 
 and $\Omega$.
 Thus, properties of M\"{o}bius functions can be used to further understand this formalism~\cite{Smatrix-singularQM}.
 By contrast,
 in this paper, we reverse the logic and
 focus on the basic principles that ultimately generate this remarkable analytic structure of the 
S-matrix for singular potentials.

The central results we address below are based on
the analytic structure of the S-matrix that relies
on Blaschke factorization. For our current purposes, some language, properties, and theorems of
Blaschke  products are in order.
Let 
$\mathbb{D} = \left\{ \Omega \in \mathbb{C} :  \; \; |\Omega| < 1 \right\}$ 
and
 $\overline{\mathbb{D}}= \left\{ \Omega \in \mathbb{C} :  \; \; |\Omega| \leq 1 \right\}$ 
be the open and closed unit disks in the complex plane, respectively;
and the boundary
$\mathbb{T} = \left\{ \Omega \in \mathbb{C} :  \; \; |\Omega| = 1 \right\} = \partial 
\overline{\mathbb{D}}
$ 
be the unit circle.
Then~\cite{Blaschke1, Blaschke2},
\begin{quotation}
\noindent
Let $F(z)$ be  a holomorphic function on 
$\mathbb{D}$
that can be extended to a continuous function 
on $\overline{\mathbb{D}}$.
 If $F$
is a mapping 
 of the unit disk
 $\overline{\mathbb{D}}$
  to itself
  that preserves the disk boundary $\mathbb{T}$
   (i.e., $|F| = 1$ if $|z| = 1 $),
 then $F$ admits a finite Blaschke product factorization,
   \begin{equation}
 F(z) = 
 \zeta \;
\prod_{j=1}^{n} 
\left(
\frac{z-z_{j}}{ 1 - z_{j}^{*} \, z}
\right)
 \; ,
 \label{eq:Blaschke_factorization}
 \end{equation}
where $\zeta$ is a phase factor ($|\zeta| =1$) independent of $z$ and $z_{j}$ are the zeros of $F(z)$
in $\mathbb{D}$.
  \end{quotation}
 For the case when a bounded analytic function satisfies the conditions above on the open disk $\mathbb{D}$, 
 Carath\'{e}odory's theorem yields a possibly infinite factorization with an appropriate behavior of the given sequence 
 of zeros~\cite{Garnett:1981}.
 In the finite case above,
 the number $n$ is the degree, ${\rm deg} \, F$, of the mapping.

Specifically, for  our generic treatment of SQM, the mapping 
$ \hat{S}   (\Omega)$
 is indeed 
  restricted to the closed unit disk, $\overline{\mathbb{D}}$,
 i.e.,  $|S| \leq 1$ iff $|\Omega| \leq 1 $.
This property can be established directly from ``conserved currents'' (with 
the usual probabilistic interpretation in
the particular case of quantum mechanics proper applications)
as described by Wronskian properties of pairs of solutions of the governing differential equation.
In effect,
from the Wronskian $W[\psi_{1},\psi_{2}]$ 
of any two solutions
of Eq.~(\ref{eq:singular-QM_radial})
 (and/or their complex conjugates, which are 
also solutions), and 
through the definition $J[u]=W[u^{*},u]/i$ of conserved currents,
it follows that $J[u_{\pm}] = \pm 2$ and similarly $J[u_{1,2}]= \pm 2$
(guaranteed by the WKB normalization).
More generally, the form 
$J[u,v]=W[u^{*}, v]/i$
is  conjugate-symmetric sesquilinear,
$J[u,\lambda_{1} v_{1} + \lambda_{2} v_{2}] 
= 
\lambda_{1} J[u, v_{1}]
+
\lambda_{2} J[u, v_{2}]
$
and
$\left( J[u,v] \right)^{*}= J[v,u]$,
and further satisfies
$\left( J[u,v] \right)^{*}= -J[u^{*},v^{*}]$.
Then,
from these definitions,
one can derive all the relevant current-conservation relations,
including explicit identities for reflection and transmission coefficients.
Moreover, the Hermitian quadratic form $J[u]\equiv J[u,u]$ is an SU(1,1) inner product in the specific sense that
$J[u^{*}]=- J[u]$, leading to 
 $J[\lambda_{1} u + \lambda_{2} u^{*}] = \left( |\lambda_{1}|^{2} -  |\lambda_{2}|^{2} \right) \, J[u]$;
 thus, 
 for wave function~(\ref{eq:wave-function_bases}) that is the general solution to 
 Eq.~(\ref{eq:singular-QM_radial}), 
 \begin{equation}
\frac{1}{2} \,
 J [u ] =
 |C^{ (1)} |^{2} 
 - 
 |C^{ (2)} |^{2}  
 =
|C^{ (+)} |^{2} 
-
|C^{ (-)} |^{2} 
\; .
\end{equation}
This implies that
 $
\bigl|  \hat{S}  \bigr|^{2} -1
=
\left[
\left| \Omega \right|^{2} -1
\right]
\,
|C^{(-)}/C^{(2)}|^{2}
$, whence
\begin{equation}
{\rm sgn}
\left( J \right)
=
{\rm sgn}
\left[
\bigl|  \hat{S}  \bigr|^{2} -1
\right]
=
{\rm sgn}
\left[
\left| \Omega \right|^{2} -1
\right]
\; .
\label{eq:S-matrix_beta_sign_connection}
\end{equation}
Therefore, 
$\bigl|  \hat{S} \bigr| \leq 1$
iff
$\left| \Omega \right| \leq 1$
(with one-to-one correspondence of the equal signs), so that
from the ensuing  map 
 $ \hat{S}  (\Omega)$  
 one concludes that 
 $ \hat{S}  (\Omega)$ 
admits the Blaschke factorization~(\ref{eq:Blaschke_factorization}),
with $z \equiv \Omega$ and $F \equiv \hat{S}$,
 up to a global phase.

The characterization of the 
form of the asymptotic S-matrix concludes with the restriction to
a  single Blaschke factor, which is due to the existence of 
a unique zero $\Omega_{1} = {\mathcal R}^{*}$ for the S-matrix
(except for $|\mathcal{R}|=1$; see below).
This property can be established from general arguments,
as follows.
First,
from the generic framework of Sec.~\ref{sec:SQM-multichannel-setup},
one  can depict the zeros and poles of the S-matrix using Fig.~1.
In effect, the zero of $\hat{S}$ occurs when the building block $u_{1}$ is suppressed. 
Second, 
define the solutions $\check{u}_{1}$ and $\check{u}_{-}$, 
with modified normalizations adapted to the standard 1D scattering problem,
\begin{eqnarray}
\check{u}_{1} 
& = &
\left\{
\begin{array}{ll}
u_{+} + {\mathcal R} u_{-}
\;  &  \; \; {\rm for} \; r \sim 0
\\
 {\mathcal T} u_{1}
 \;  &  \; \; {\rm for} \; r \sim \infty
\end{array}
\right.
\label{eq:tilde-u_1}
\; ,
\\
\check{u}_{-}
& = &
\left\{
\begin{array}{ll}
u_{2} + {\mathcal R'} u_{1}
\;  &  \; \; {\rm for} \; r \sim \infty
\\
 {\mathcal T'} u_{-}
 \; &  \; \; {\rm for} \; r \sim 0
 \; 
\end{array}
\label{eq:tilde-u_-}
\; ,
\right.
\end{eqnarray} 
where 
$( {\mathcal R}  , {\mathcal T}) $ 
and
$( {\mathcal R}'  , {\mathcal T}') $ 
are,
respectively,
the right-moving and left-moving reflection and transmission amplitude coefficients.
Third, by time-reversal invariance
(given by the complex conjugate and reversal of the arrows in Fig.~1),
we can see that this amounts to
$
\check{u}_{1}^{*} 
= \left( {\mathcal T} u_{1} \right)^{*}
=  {\mathcal T}^{*} u_{2}$,
where Eq.~(\ref{eq:conjugate-basis-12-relation})
was used---this effectively
suppresses $u_{1}$ and selects the zero of $\hat{S}$;
 thus,
$\check{u}_{1}^{*}  
=
[ u_{+} ]^{*} + {\mathcal R}^{*} [ u_{-} ]^{*}
=
 u_{-}  + {\mathcal R}^{*}  u_{+} $,
leading to
\begin{equation}
\Omega_{1} = \left. \Omega \right|_{\rm zero \; of \; \hat{S} } = {\mathcal R}^{*}
\; 
\end{equation}
[by comparison against Eq.~(\ref{eq:wave-function_near-origin-basis})].
Fourth, from the physical bound $|{\mathcal R}| \leq 1 $,
which is established through $|{\mathcal R}|^2 + |{\mathcal T}|^2 = 1$ via Wronskian relations (see below),
it follows that
\begin{equation}
 \text{\em 
 for $|{\mathcal R}| < 1 $,
 there is a unique zero $\Omega_{1} = {\mathcal R}^{*}$ of the S-matrix, 
 with $\Omega_{1} \in
  {\mathbb{D}} $.}
 \label{eq:unique-zero}
 \end{equation}
 So far, we have only established that
  $\Omega_{1} \in
  \overline{\mathbb{D}} $, but a zero on the unit circle
  $\mathbb{T} = \partial  \overline{\mathbb{D}} $ is to be rejected,
  as shown below.
Fifth, by comparison against 
Eq.~(\ref{eq:Blaschke_factorization}), one can explicitly write the reduced S-matrix as a single Blaschke 
factor times a global phase,
\begin{equation}
\hat{S}
=
\Delta 
\,
\frac{
  \Omega
-
 {\mathcal R}^{*}
}{
{\mathcal R}
\,
\, \Omega
-
1
}
\label{eq:asymptotic-S-matrix_from-MC-S-matrix}
\; .
\end{equation}
Sixth,
regarding the general Blaschke-factor
of Eq.~(\ref{eq:asymptotic-S-matrix_from-MC-S-matrix}),
when $|\mathcal{R}| =1$,
the S-matrix is trivial with respect to $\Omega$,
in the sense that it is restricted to the unit circle, $|\hat{S}| =1$ and $\Omega$-independent,
as can be easily verified
(this is a general property of the Blaschke factors);
thus, in that case,
being a constant, 
$\hat{S}$ has neither zeros nor poles---incidentally,
there is no contradiction herein because
$\mathcal{T}=0$ when  $|\mathcal{R}| =1$, and the asymptotic form of
Eq.~(\ref{eq:tilde-u_1})
becomes degenerate, 
failing to produce an actual zero of $\hat{S}$.
Finally, in Eq.~(\ref{eq:asymptotic-S-matrix_from-MC-S-matrix}), it is also possible to identify the global phase 
as
\begin{equation}
\Delta =  
 - \frac{ {\mathcal T} }{ {\mathcal T}^{*}}
  =
  \frac{ {\mathcal R'} }{ {\mathcal R}^{*}}
  \label{eq:S-matrix_Mobius-phase}
  \; ,
\end{equation}
from the following general arguments:
(i)
$\Delta ={\mathcal R'}/ {\mathcal R}^{*}$
from $\hat{S} (\Omega =0) = {\mathcal R'}$
(i.e., the corresponding S-matrix is the left-moving reflection coefficient);
(ii) by means of a Wronskian relation,
the Stokes' reciprocity relation 
$  {\mathcal T} {\mathcal R}^{*}
+
{\mathcal T}^{*}
     {\mathcal R'} 
= 0
$
is established.
Parenthetically,
a network of Wronskian relations 
$W \bigl[  \check{u}_{1}^{*}, \check{u}_{1} \bigr]$,
$W \bigl[  \check{u}_{-}^{*}, \check{u}_{-} \bigr]$,
$W \bigl[  \check{u}_{1}^{*}, \check{u}_{-} \bigr]$,
and
$W \bigl[  \check{u}_{1}, \check{u}_{-} \bigr]$,
implies the following four conditions:
 $|{\mathcal R}|^{2}  +  |{\mathcal T}|^{2} = 1
 $,
 $|{\mathcal R'}|^{2}  +  |{\mathcal T'}|^{2} = 1$,
${\mathcal R^{*} } {\mathcal T'} + {\mathcal T^{*} } {\mathcal R'} = 0$,
and
${\mathcal T'} = {\mathcal T}$
(with the last two being used in the above argument).

It should be noticed that various particular cases of
Eq.~(\ref{eq:asymptotic-S-matrix_from-MC-S-matrix})
have appeared in the 
literature in Refs.~\cite{perelomov:70,Alliluev_ISP,aud:99}
(for $\Omega =0$)
and in Ref.~\cite{Voronin}
(for $\Omega =0$ and  $\Omega =\infty$),
while the general form 
has been derived by other techniques in Refs.~\cite{Esposito,Stroffolini,Smatrix-singularQM}.

As a corollary of relations~(\ref{eq:unique-zero})--(\ref{eq:S-matrix_Mobius-phase}),
\begin{quotation}
\noindent
{\em  the  function 
$\hat{S} (\Omega)$ 
is {\em analytic\/} in the closed unit circle
$\overline{\mathbb{D}} $,
i.e., it is a meromorphic
function with no poles for $| \Omega | \leq 1$.}
 \end{quotation}
This is simply due to the fact that the unique purported pole is located at $\Omega_{2} = 1/{\mathcal R}$, 
as shown by the Blaschke factor---but this location of $\Omega_{2}$
   could also be established independently
by a similar subset of arguments, 
via the suppression of the building block $u_{2}$ 
combined with
Eq.~(\ref{eq:tilde-u_1}) leading to
$\check{u}_{1}
=
 u_{+} + {\mathcal R}  u_{-} $
 (i.e, 
 $\Omega$ given by the explicit ratio $ \Omega_{2}  = 1/\mathcal{R} $).
Thus, by
a reversal of the reflection-coefficient bound discussed above,
$|\Omega_{2}| = |1/{\mathcal R}| = 1/|\Omega_{1}| \geq 1$;
in addition,
 the
trivial case $|{\mathcal R}| =1 $
 can be dealt with separately 
 [as established in the argument following
 Eq.~(\ref{eq:asymptotic-S-matrix_from-MC-S-matrix})],
 thus leading to  $\hat{S}$ being a constant ($\Omega$-independent) of modulus one,
 so that the stricter condition $|\Omega_{2}| > 1$ is enforced.
 
Incidentally, the S-matrix of
Eq.~(\ref{eq:asymptotic-S-matrix_from-MC-S-matrix}),
being a single Blaschke factor, 
defines a well-known M\"{o}bius transformation 
$\hat{S} (\Omega)$
that is an automorphism of the unit disk~\cite{Needham_Complex}.

\section{Determination of Nonunitary Solutions from the Whole Family
of Unitary Solutions}
\label{sec:unitary-nonunitary_connection}

The general properties
 discussed in Sec.~\ref{sec:S-matrix-structure}
 for the
 two-channel case (with one singularity other than asymptotic infinity)
 yield  a holomorphic function
$ \hat{S}  (\Omega)$ 
 with a unique zero at $\Omega_{1} = \mathcal{R}^{*}$.
Furthermore,
the scattering bound $|{\mathcal R}| \leq 1$
(familiar restriction on the reflection coefficients)
leads to a location of 
the corresponding pole outside the unit disk $\mathbb{D}$, 
so that the function $ \hat{S}  (\Omega)$ 
is guaranteed to be analytic  in
$ \overline{\mathbb{D}} $ (i.e., $|\Omega | \leq 1$).

As a result,
Cauchy's integral formula~\cite{Needham_Complex}
 implies the following 
{\em ``average characterization''
of nonunitary solutions for all the cases with absorption due to a singular potential\/},
\begin{equation}
\hat{S}(\Omega) 
= 
\frac{1}{2 \pi i }
\oint\limits_{{\mathbb{T}} }
d \Omega' \, \frac{ \hat{S} (\Omega') }{\Omega'-\Omega } 
\; ,
\label{eq:Cauchy-theorem}
\end{equation}
where $\mathbb{T} = \partial \overline{\mathbb{D}} $ is the unit circle consisting of 
the whole family of unitary S-matrix 
values,
 i.e., the whole family of 
solutions to Eq.~(\ref{eq:singular-QM_radial})
with 
 ``elastic'' or self-adjoint boundary conditions.
  This amounts to 
 $\Omega' = e^{i\chi} $, where $\chi \in \mathbb{R}$ is a phase specifying the generic self-adjoint condition.
The one-to-one correspondence of the conditions
$|\Omega' | = 1$
and 
$|S (\Omega' )| = |\hat{S} (\Omega' ) | = 1 $
can be seen from Eq.~(\ref{eq:S-matrix_beta_sign_connection}).
With this notation,
and
going back to the primitive form of  asymptotic S-matrix~(\ref{eq:S-matrix_and_reduced-S-matrix}) by reintroducing the 
required phase factors (whenever appropriate),
Eq.~(\ref{eq:Cauchy-theorem}) 
 can be rewritten in the suggestive form 
\begin{equation}
{S}(\Omega) 
= 
\int_{0}^{2\pi}
\frac{d \chi}{2 \pi  }
\; 
 \frac{ {S}(\Omega' = e^{i\chi})  }{ 1 -\Omega \, e^{-i\chi} }
\; .
\label{eq:Cauchy-average}
\end{equation}
Equation~(\ref{eq:Cauchy-average})
 is an ensemble average with a nonuniform (shifted) weight of the whole family of unitary S-matrices.
Therefore,
we have established the general identity relating unitary and nonunitary solutions
 for $|\Omega |< 1$, i.e., all the cases that exhibit 
{\em net absorption\/}.

The particular case of ``perfect absorption,'' $\Omega = 0$,
involves a symmetric ensemble average with uniform weight, i.e.,
\begin{equation}
{S} (\Omega=0) 
= 
\int_{0}^{2\pi}
\frac{d \chi}{2 \pi  }
\; 
 {S} (\Omega' = e^{i\chi}) 
\; ,
\end{equation}
which corresponds to the particular case
 considered in Refs.~\cite{radin:75} and \cite{bawin_coon_wire}.
It is also noteworthy that this is equal to the left-moving reflection coefficient:
$\hat{S} (\Omega=0) = {\mathcal R}'$
[see Eqs.~(\ref{eq:asymptotic-S-matrix_from-MC-S-matrix}) and (\ref{eq:S-matrix_Mobius-phase})].
Our present work shows that this result is not accidental 
but the logical consequence of general physical principles applied to the 
S-matrix---following a similar methodology to the S-matrix approach in field theory.
Remarkably, this is a generic property of SQM (either conformal or an irregular singularity):
while singular potentials support this analytic ``structure,''
the distinct case (not tackled herein) of absorption by complex (optical) potentials generally
does not.

\section{Conclusions}
\label{sec:conclusions}

We have shown that,
given the existence of a singular problem (in the  sense of SQM),
 a basic set of general principles:
 linearity,
unitarity of the multichannel S-matrix with the related network of current-conservation statements
(encoded by reflection and transmission coefficients),
and time-reversal symmetry allow the complete determination of the analytic structure of the asymptotic S-matrix
of SQM with respect to the parameter $\Omega$ that specifies boundary conditions.
This analytic structure involves M\"{o}bius transformations of the Blaschke-factor type associated with the unit disk.
Due to its generality, this approach also provides the rationale to predict and explain 
 the intriguing average relationship between unitary and nonunitary solutions at the level of the asymptotic S-matrix.

\acknowledgements
This work was partially supported by
the University of San Francisco Faculty Development Fund
(H.E.C.); and ANPCyT, Argentina (L.N.E., H.F., and C.A.G.C.).

\newpage

\begin{figure}[ht]
\centering
\resizebox{6.5in}{!}{\includegraphics{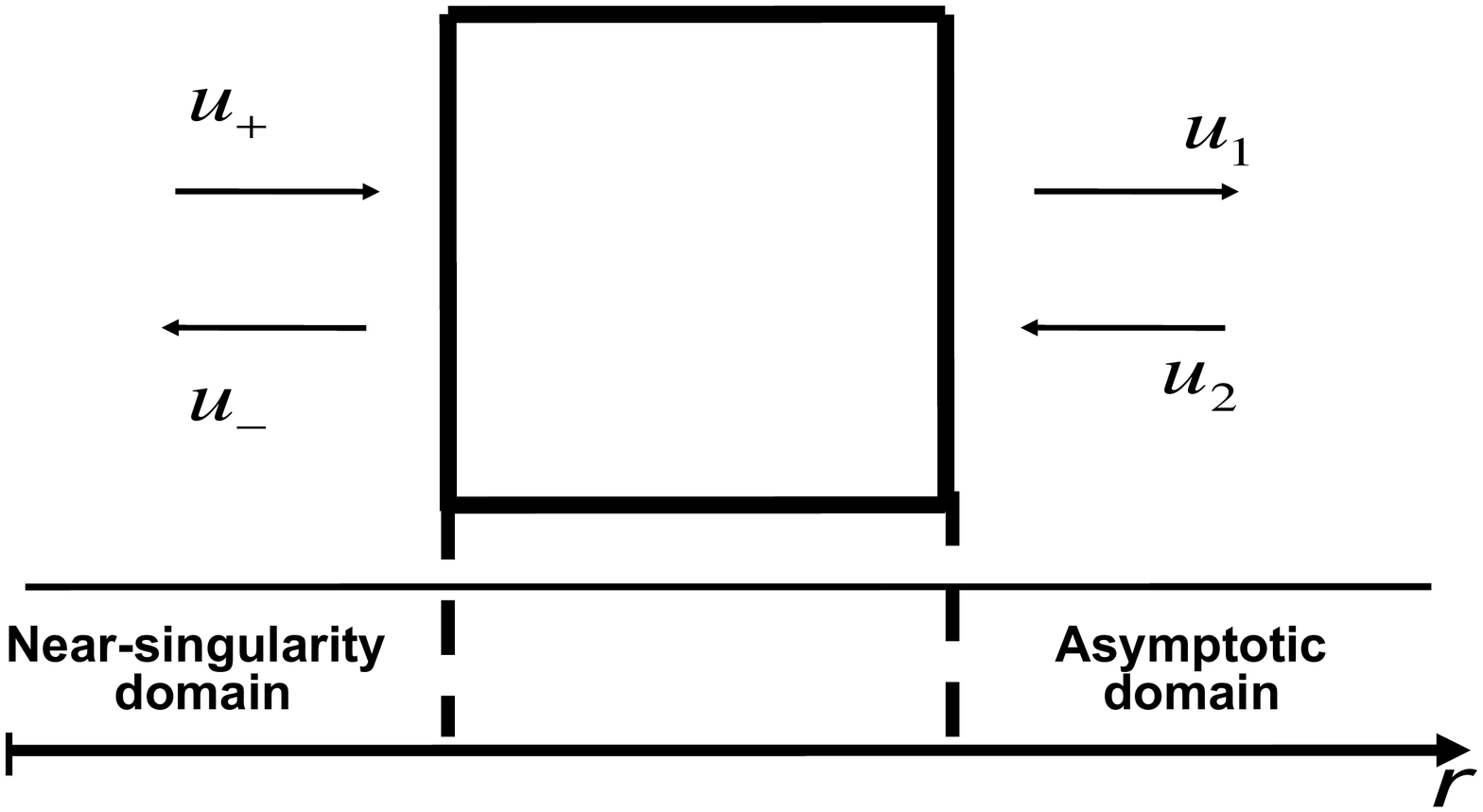}}
\caption{The multichannel framework can be visualized 
as a connector between  two domains, each involving a singular point:
the asymptotic domain ($r \sim \infty$), with outgoing/ingoing basis functions 
${\bf u_{1}}$ and ${\bf u_{2}}$, and the
  near-singularity domain
($r \sim 0$),
with outgoing/ingoing basis functions 
${\bf u_{+}}$ and ${\bf u_{-}}$.}
\label{fig:multichannel-analytic}
\end{figure}

\end{document}